\documentclass[aps,prd,superscriptaddress,showpacs,twocolumn]{revtex4}

\usepackage{graphicx}
\usepackage{epsfig}
\usepackage{dcolumn}

\newcommand{\gevcs}{\hbox{GeV}/c^2}
\newcommand{\gev}{\hbox{GeV}}
\newcommand{\mevcs}{\hbox{MeV}/c^2}
\newcommand{\mev}{\hbox{MeV}}

\newcommand{\MMS}{M_{\rm rec}^2}
\newcommand{\yones}{\Upsilon(1S)}
\newcommand{\ytwos}{\Upsilon(2S)}
\newcommand{\gR}{\gamma_{\rm R}}
\newcommand{\gisr}{\gamma_{\rm ISR}}
\newcommand{\x}{X(3872)}

\newcommand{\cdfy}{Y(4140)}

\newcommand{\LK}{\mathcal{L}}
\newcommand{\eff}{\varepsilon}
\newcommand{\BR}{{\cal B}}

\newcommand{\piz}{\pi^0}

\newcommand{\etac}{\eta_c}
\newcommand{\chicz}{\chi_{c0}}
\newcommand{\chico}{\chi_{c1}}
\newcommand{\chict}{\chi_{c2}}
\newcommand{\chicJ}{\chi_{cJ}}

\newcommand{\psp}{\psi(2S)}

\newcommand{\jpsi}{J/\psi}

\newcommand{\EE}{e^+e^-}
\newcommand{\MM}{\mu^+\mu^-}

\newcommand{\pp}{\pi^+\pi^-}
\newcommand{\kk}{K^+K^-}

\newcommand{\kkpi}{K_S^0K^+\pi^- + c.c.}
\newcommand{\ppjpsi}{\pi^+\pi^- J/\psi}

\newcommand{\beq}{\begin{equation}}
\newcommand{\eeq}{\end{equation}}
\newcommand{\bitm}{\begin{itemize}}
\newcommand{\eitm}{\end{itemize}}
\def\Journal#1#2#3#4{{#1} {\bf #2}, #3 (#4)}

\def\PRL{Phys. Rev. Lett.}
\def\PRD{Phys. Rev. D}

\begin{document}

%************************************************************
\preprint{} \preprint{\vbox{ \hbox{   }
%  \hbox{Belle DRAFT {\it 10-02 [20-June-10]}}
  \hbox{Intended for {\it Phys. Rev. D(R)}}
  \hbox{Authors: X.~L.~Wang, C. P. Shen C.~Z.~Yuan, and P.~Wang }
  \hbox{Committee: S. Uehara (chair), S. Lange, H. Nakazawa}
                        \hbox{Belle Preprint 2011-12}
                        \hbox{KEK   Preprint 2011-10}
%                        \hbox{BIHEP-EP-2010-001}
                        }}
\title{
%%\quad\\[1.4cm]
Search for charmonium and charmonium-like states in $\Upsilon(2S)$
radiative decays}

%%%%  author list %%%%%%%
%\input{pub349}
%%%\affiliation{University of Bonn, Bonn}
\affiliation{Budker Institute of Nuclear Physics SB RAS and Novosibirsk State University, Novosibirsk 630090}
\affiliation{Faculty of Mathematics and Physics, Charles University, Prague}
%%%\affiliation{Chiba University, Chiba}
\affiliation{University of Cincinnati, Cincinnati, Ohio 45221}
\affiliation{Department of Physics, Fu Jen Catholic University, Taipei}
\affiliation{Justus-Liebig-Universit\"at Gie\ss{}en, Gie\ss{}en}
\affiliation{Gifu University, Gifu}
%%%\affiliation{The Graduate University for Advanced Studies, Hayama}
%%%\affiliation{Gyeongsang National University, Chinju}
\affiliation{Hanyang University, Seoul}
\affiliation{University of Hawaii, Honolulu, Hawaii 96822}
\affiliation{High Energy Accelerator Research Organization (KEK), Tsukuba}
\affiliation{Hiroshima Institute of Technology, Hiroshima}
%%%\affiliation{University of Illinois at Urbana-Champaign, Urbana, Illinois 61801}
\affiliation{Indian Institute of Technology Guwahati, Guwahati}
\affiliation{Indian Institute of Technology Madras, Madras}
%%%\affiliation{Indiana University, Bloomington, Indiana 47408}
\affiliation{Institute of High Energy Physics, Chinese Academy of Sciences, Beijing}
\affiliation{Institute of High Energy Physics, Vienna}
\affiliation{Institute of High Energy Physics, Protvino}
%%%\affiliation{Institute of Mathematical Sciences, Chennai}
\affiliation{INFN - Sezione di Torino, Torino}
\affiliation{Institute for Theoretical and Experimental Physics, Moscow}
\affiliation{J. Stefan Institute, Ljubljana}
\affiliation{Kanagawa University, Yokohama}
\affiliation{Institut f\"ur Experimentelle Kernphysik, Karlsruher Institut f\"ur Technologie, Karlsruhe}
\affiliation{Korea Institute of Science and Technology Information, Daejeon}
\affiliation{Korea University, Seoul}
%%%\affiliation{Kyoto University, Kyoto}
\affiliation{Kyungpook National University, Taegu}
\affiliation{\'Ecole Polytechnique F\'ed\'erale de Lausanne (EPFL), Lausanne}
\affiliation{Faculty of Mathematics and Physics, University of Ljubljana, Ljubljana}
\affiliation{Luther College, Decorah, Iowa 52101}
\affiliation{University of Maribor, Maribor}
\affiliation{Max-Planck-Institut f\"ur Physik, M\"unchen}
\affiliation{University of Melbourne, School of Physics, Victoria 3010}
\affiliation{Nagoya University, Nagoya}
%%%\affiliation{Nara University of Education, Nara}
\affiliation{Nara Women's University, Nara}
\affiliation{National Central University, Chung-li}
\affiliation{National United University, Miao Li}
\affiliation{Department of Physics, National Taiwan University, Taipei}
\affiliation{H. Niewodniczanski Institute of Nuclear Physics, Krakow}
\affiliation{Nippon Dental University, Niigata}
\affiliation{Niigata University, Niigata}
\affiliation{University of Nova Gorica, Nova Gorica}
\affiliation{Osaka City University, Osaka}
%%%\affiliation{Osaka University, Osaka}
\affiliation{Pacific Northwest National Laboratory, Richland, Washington 99352}
%%%\affiliation{Panjab University, Chandigarh}
%%%\affiliation{Peking University, Beijing}
%%%\affiliation{Princeton University, Princeton, New Jersey 08544}
\affiliation{Research Center for Nuclear Physics, Osaka}
%%%\affiliation{RIKEN BNL Research Center, Upton, New York 11973}
%%%\affiliation{Saga University, Saga}
\affiliation{University of Science and Technology of China, Hefei}
\affiliation{Seoul National University, Seoul}
%%%\affiliation{Shinshu University, Nagano}
\affiliation{Sungkyunkwan University, Suwon}
\affiliation{School of Physics, University of Sydney, NSW 2006}
\affiliation{Tata Institute of Fundamental Research, Mumbai}
\affiliation{Excellence Cluster Universe, Technische Universit\"at M\"unchen, Garching}
\affiliation{Toho University, Funabashi}
\affiliation{Tohoku Gakuin University, Tagajo}
\affiliation{Tohoku University, Sendai}
\affiliation{Department of Physics, University of Tokyo, Tokyo}
\affiliation{Tokyo Institute of Technology, Tokyo}
%%%\affiliation{Tokyo Metropolitan University, Tokyo}
\affiliation{Tokyo University of Agriculture and Technology, Tokyo}
%%%\affiliation{Toyama National College of Maritime Technology, Toyama}
\affiliation{CNP, Virginia Polytechnic Institute and State University, Blacksburg, Virginia 24061}
%%%\affiliation{Wayne State University, Detroit, Michigan 48202}
%%%\affiliation{Yamagata University, Yamagata}
\affiliation{Yonsei University, Seoul}
  \author{X.~L.~Wang}\affiliation{Institute of High Energy Physics, Chinese Academy of Sciences, Beijing} % IHEP
  \author{C.~P.~Shen}\affiliation{Nagoya University, Nagoya} % Nagoya
  \author{C.~Z.~Yuan}\affiliation{Institute of High Energy Physics, Chinese Academy of Sciences, Beijing} % IHEP
  \author{P.~Wang}\affiliation{Institute of High Energy Physics, Chinese Academy of Sciences, Beijing} % IHEP
  \author{I.~Adachi}\affiliation{High Energy Accelerator Research Organization (KEK), Tsukuba} % KEK
% \author{K.~Adamczyk}\affiliation{H. Niewodniczanski Institute of Nuclear Physics, Krakow} % Krakow
  \author{H.~Aihara}\affiliation{Department of Physics, University of Tokyo, Tokyo} % Tokyo
% \author{K.~Arinstein}\affiliation{Budker Institute of Nuclear Physics SB RAS and Novosibirsk State University, Novosibirsk 630090} % BINP
% \author{Y.~Arita}\affiliation{Nagoya University, Nagoya} % Nagoya
  \author{D.~M.~Asner}\affiliation{Pacific Northwest National Laboratory, Richland, Washington 99352} % PNNL
% \author{T.~Aso}\affiliation{Toyama National College of Maritime Technology, Toyama} % Toyama
% \author{V.~Aulchenko}\affiliation{Budker Institute of Nuclear Physics SB RAS and Novosibirsk State University, Novosibirsk 630090} % BINP
 \author{T.~Aushev}\affiliation{Institute for Theoretical and Experimental Physics, Moscow} % ITEP
% \author{T.~Aziz}\affiliation{Tata Institute of Fundamental Research, Mumbai} % Tata
  \author{A.~M.~Bakich}\affiliation{School of Physics, University of Sydney, NSW 2006} % Sydney
% \author{Y.~Ban}\affiliation{Peking University, Beijing} % Peking
  \author{E.~Barberio}\affiliation{University of Melbourne, School of Physics, Victoria 3010} % Melbourne
% \author{A.~Bay}\affiliation{\'Ecole Polytechnique F\'ed\'erale de Lausanne (EPFL), Lausanne} % Lausanne
% \author{I.~Bedny}\affiliation{Budker Institute of Nuclear Physics SB RAS and Novosibirsk State University, Novosibirsk 630090} % BINP
% \author{M.~Belhorn}\affiliation{University of Cincinnati, Cincinnati, Ohio 45221} % Cincinnati
% ...
% BINP
% \author{P.~Zhou}\affiliation{Wayne State University, Detroit, Michigan 48202} % WayneState
% \author{V.~Zhulanov}\affiliation{Budker Institute of Nuclear Physics SB RAS and Novosibirsk State University, Novosibirsk 630090} % BINP
% \author{T.~Zivko}\affiliation{J. Stefan Institute, Ljubljana} % Ljubljana
% \author{A.~Zupanc}\affiliation{Institut f\"ur Experimentelle Kernphysik, Karlsruher Institut f\"ur Technologie, Karlsruhe} % Karlsruhe
% \author{N.~Zwahlen}\affiliation{\'Ecole Polytechnique F\'ed\'erale de Lausanne (EPFL), Lausanne} % Lausanne
% \author{O.~Zyukova}\affiliation{Budker Institute of Nuclear Physics SB RAS and Novosibirsk State University, Novosibirsk 630090} % BINP
\collaboration{The Belle Collaboration}

\date{\today}

\begin{abstract}

Using a sample of 158~million $\ytwos$ events collected with the
Belle detector, charmonium and charmonium-like states with even
charge parity are searched for in $\ytwos$ radiative decays. No
significant $\chicJ$ or $\etac$ signal is observed and  the
following upper limits at 90\% confidence level (C.L.) are obtained: $\BR(\ytwos\to
\gamma \chi_{c0})< 1.0 \times 10^{-4}$, $\BR(\ytwos\to \gamma
\chico)<3.6 \times 10^{-6}$, $\BR(\ytwos\to \gamma \chi_{c2})<1.5
\times 10^{-5}$, and $\BR(\ytwos\to \gamma \etac)< 2.7 \times
10^{-5}$. No significant signal of any charmonium-like state is
observed, and we obtain the limits $\BR(\ytwos\to \gamma
\x)\times\BR(\x\to\pi^+\pi^-\jpsi)< 0.8 \times 10^{-6}$,
$\BR(\ytwos \to \gamma \x)\times\BR(\x\to\pi^+\pi^-\pi^0 \jpsi)<
2.4\times 10^{-6}$, $\BR(\ytwos \to \gamma
X(3915))\times\BR(X(3915)\to\omega \jpsi)< 2.8\times 10^{-6}$,
$\BR(\ytwos \to \gamma Y(4140))\times\BR(Y(4140)\to\phi\jpsi)) <
1.2\times 10^{-6}$, and $\BR(\ytwos \to \gamma
X(4350))\times\BR(X(4350)\to\phi\jpsi))< 1.3\times 10^{-6}$ at
90\% C.L.
% Furthermore, no evidence is found for
%excited charmonium states below $5.5~\gevcs$ in the invariant
%mass distributions of any of the final states.

\end{abstract}


\maketitle

%% introduction

The data samples of the $B$ factories have provided a wealth of
experimental information on charmonium spectroscopy~\cite{review}.
Below open charm threshold agreement between experimental mass
measurements and predictions based upon potential models was
recently demonstrated with high accuracy for the
$h_c$~\cite{hc1,hc2}. However, in the region above the open charm
threshold, in addition to many conventional charmonium states, a
number of charmonium-like states (the so-called ``$XYZ$
particles") have been discovered with unusual properties. These
may include exotic states, such as quark-gluon hybrids, meson
molecules, and multi-quark states~\cite{review}. Many of these new
states are established in a single production mechanism or in a
single decay mode only. To better understand them, it is necessary
to search for such states in more production processes and/or
decay modes. States with $J^{PC}=1^{--}$ can be studied via
initial state radiation (ISR) with the large $\Upsilon(4S)$ data
samples at BaBar or Belle, or via $\EE$ collisions directly at the
peak energy at, for example, BESIII. For charge-parity-even
charmonium states, radiative decays of the narrow $\Upsilon$
states below the open bottom threshold can be examined.

The production rates of the $P$-wave spin-triplet $\chicJ$~($J$=0,
1, 2) and $S$-wave spin-singlet $\etac$ states in $\yones$
radiative decays have been calculated by Gao {\it et al.}; the
rates in $\ytwos$ decays are estimated to be at the same
level~\cite{ktchao}. However, there are no such calculations or
estimations for ``$XYZ$ particles" due to the limited knowledge of
their nature.

In this paper, with the world largest data sample taken at the
$\ytwos$ peak, we report a search for the $\chicJ$, $\eta_c$,
$\x$~\cite{bellex}, $X(3915)$~\cite{uehara}, and
$\cdfy$~\cite{cdfy} in $\ytwos$ radiative decays, extending our
previous work on the $\yones$ sample~\cite{yonesRD}. In addition,
the new structure $X(4350)$~\cite{bellex4350}, which was observed
as a 3.2 standard deviation ($\sigma$) signal in $\gamma\gamma\to \phi\jpsi$ is also
searched for. As any charmonium state above $\psp$ is expected to
have a larger branching fraction for the E1/M1 transition to
$\psp$ than to $\jpsi$~\cite{Barnes}, we also search for states
decaying into $\gamma\psp$.
%The $\chicJ$ states are reconstructed
%via their E1 transitions to the $\jpsi$. The $\etac$ is
%reconstructed in the $\kkpi$, $\pp\kk$, $2(\kk)$, $2(\pp)$, and
%$3(\pp)$ final states. To search for the $\x$ and $X(3915)$, we
%use the $\ppjpsi$ and $\pp\piz\jpsi$ final states, while the
%$\cdfy$ and $X(4350)$ are searched for in the $\phi\jpsi$ mode.

The data used in this analysis include a 24.7~fb$^{-1}$ data
sample collected at the $\ytwos$ peak and a 1.7~fb$^{-1}$ data
sample collected at $\sqrt{s}=9.993$~GeV (off-resonance data) with
the Belle detector~\cite{Belle} operating at the KEKB
asymmetric-energy $\EE$ collider~\cite{KEKB}. The number of the
$\ytwos$ events is determined by counting the hadronic events in
the data taken at the $\ytwos$ peak after subtracting the scaled
continuum background from the data sample collected at $\sqrt{s} =
9.993~\gev$. The selection criteria for hadronic events are
validated with the off-resonance data by comparing the measured
$R$ value ($R=\frac{\sigma(\EE\to hadrons)}{\sigma(\EE\to \MM)}$)
with CLEO's result~\cite{cleoR}. The number of $\ytwos$ events is
determined to be $(158\pm 4)\times 10^6$, with the error dominated
by the MC simulation of the $\ytwos$ decay dynamics using {\sc
pythia}~\cite{pythia}.

Well measured charged tracks and photon candidates are first
selected. For a charged track, the impact parameters perpendicular
to and along the beam direction with respect to the interaction
point (IP) are required to be less than 0.5~cm and 4~cm,
respectively, and the transverse momentum should exceed
0.1~GeV/$c$ in the laboratory frame. Information from different
detector subsystems is combined to form a likelihood
$\mathcal{L}_i$ for each particle species~\cite{pid}. A track with
$\mathcal{R}_K = \frac{\mathcal{L}_K} {\mathcal{L}_K +
\mathcal{L}_\pi}> 0.6$ is identified as a kaon, while a track with
$\mathcal{R}_K<0.4$ is treated as a pion. With this selection, the
kaon (pion) identification efficiency is about 90\% (96\%), while
5\% (6\%) of kaons (pions) are misidentified as pions (kaons).
For electron identification, the likelihood ratio is
defined as $\mathcal{R}_e = \frac{\mathcal{L}_e} {\mathcal{L}_e +
\mathcal{L}_x}$, where $\mathcal{L}_e$ and $\mathcal{L}_x$ are the
likelihoods for electron and non-electron, respectively,
determined using the ratio of the energy deposited in the
electromagnetic calorimeter (ECL) to the momentum measured in the
silicon vertex detector and central drift chamber (CDC), the
shower shape in the ECL, the matching between the position of
charged track trajectory and the cluster position in the ECL, the
hit information from the aerogel threshold Cherenkov counters and
the dE/dx measurements in the CDC~\cite{EID}. For muon
identification, the likelihood ratio is defined as
$\mathcal{R}_\mu = \frac{\mathcal{L}_\mu} {\mathcal{L}_\mu +
\mathcal{L}_\pi + \mathcal{L}_K}$, where $\mathcal{L}_\mu$,
$\mathcal{L}_\pi$ and $\mathcal{L}_K$ are the likelihoods for
muon, pion and kaon hypotheses, respectively, based on the
matching quality and penetration depth of associated hits in the
iron flux return (KLM)~\cite{MUID}.
%Similar likelihood ratios are formed for electrons using
%additional information from the electromagnetic calorimeter
%(ECL)~\cite{EID} and for muons with additional information from
%RPC's (resistive plate counters) in the iron flux return
%(KLM)~\cite{MUID}.
A good neutral cluster is reconstructed as a photon if its ECL
shower does not match the extrapolation of any charged track and
its energy is greater than 40~MeV. In the $\EE$ center-of-mass
(C.M.) frame, the photon candidate with the maximum energy is
taken to be the $\ytwos$ radiative decay photon (denoted as
$\gR$), and its energy is required to be greater than
$3.5~\hbox{GeV}$. A $3.5~\hbox{GeV}$ photon energy corresponds
to a particle of mass $5.5~\gevcs$ produced in $\ytwos$ radiative
decays.

We reconstruct $\jpsi$ signals from $\EE$ or $\MM$ candidates. In
order to reduce the effect of bremsstrahlung or final-state
radiation, photons detected in the ECL within 0.05~radians of the
original $e^+$ or $e^-$ direction are included in the calculation
of the $e^+/e^-$ momentum. For the lepton pair used to reconstruct
$\jpsi$, at least one track should have $\mathcal{R}_e>0.95$ while
the other should satisfy $\mathcal{R}_e>0.05$ in the $\EE$ mode;
or one track should have $\mathcal{R}_\mu>0.95$ (in the $\chicJ$
analysis, the other track should have associated hits in the KLM
detector that agree with the extrapolated trajectory of a charged
track provided by the drift chamber) in the $\MM$ mode. The lepton
pair identification efficiency is about 97\% for $\jpsi \to e^+
e^-$ and 87\% for $\jpsi \to \mu^+ \mu^-$. In order to improve the
$\jpsi$ momentum resolution, a mass-constrainted fit is then
performed for $\jpsi$ signals in the $\gamma\jpsi$, $\ppjpsi$,
$\pp\pi^0\jpsi$, and $\phi\jpsi$ modes. Different modes have
similar $\jpsi$ mass resolutions. The $\jpsi$ signal region is
defined as $|M_{\ell^+\ell^-}-m_{\jpsi}|<30~\hbox{MeV}/c^2$
($\approx 2.5\sigma$), where $m_{\jpsi}$ is the nominal mass of
$\jpsi$. The $\jpsi$ mass sidebands are defined as
$2.959~\hbox{GeV}/c^2<M_{\ell^+\ell^-}<3.019~\hbox{GeV}/c^2$ and
$3.175~\hbox{GeV}/c^2<M_{\ell^+\ell^-}<3.235~\hbox{GeV}/c^2$, and
are twice as wide as the signal region.  For the
$\gamma\psp$ channel, the $\psp$ is reconstructed from the
$\ppjpsi$ final state, with a mass constrained to the  nominal
$\psp$ mass to improve its momentum resolution. To estimate the difference
in the $\psp$ mass resolution between MC simulation and data, the process
$\EE\to\gisr \psp$ is selected as a reference sample, and the mass
resolution is found to be $3.0\pm 0.1~\mevcs$ from data, and
$2.6~\mevcs$ from MC simulation. The difference in the mass
resolution is included when extracting the signal yields in the
analyses below.

%% data selection for chi_cJ

We search for the $\chicJ$ in the $\gamma\jpsi$ mode. The energy
deposited by the $\chicJ$ photon (denoted as $\gamma_l$, since its
energy is much lower than that of $\gR$) is required to be greater
than $150~\mev$ to reduce the large background from
mis-reconstructed photons, and the total number of photons is
required to be exactly two to suppress multiphoton backgrounds.
The angle between the $\gR$ and $\gamma_l$ should be larger than
$18^{\circ}$ to remove the background from split-off fake photons.
To remove the ISR background $\EE\to \gisr\psp\to \gisr \gamma
\chicJ$, where a photon is missing, we require the square of the
``mass recoiling against the $\gamma_l$ and $\jpsi$" ($\MMS =
(P_{\EE}-P_f)^2$, here $P_{\EE}$ is the 4-momentum of the $\EE$
collision system, and $P_f$ is the sum of the 4-momenta of the
observed final state particles) to be within $-0.5~{\rm
GeV}^2/c^4$ and 0.5~GeV$^2/c^4$. This $\MMS$ requirement is
effective since this background has at least two missing photons
and $\MMS(\gamma_l\jpsi)$ tends to be large. Bhabha and dimuon
background events with final-state radiative photons are further
suppressed by removing events in which a photon is detected within
a $18^\circ$ cone around each charged track direction.

\begin{figure}[htbp]
\psfig{file=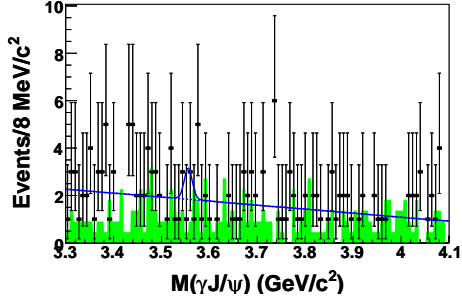,height=6cm,angle=-90} \caption{ The
$\gamma_{l}\jpsi$ invariant mass distribution in the $\ytwos$ data
sample.  There is no $\chicz$, $\chico$, or $\chict$ signal
observed. The solid curve is the best fit, the dashed curve is the
background, and the shaded histogram is from the normalized
$\jpsi$ mass sidebands. The signal yield is required to
be non-negative in the fit. } \label{mgll}
\end{figure}

The $\MM$ mode shows a clear $\jpsi$ signal, while the $\EE$ mode
has some residual radiative Bhabha background. Figure~\ref{mgll}
shows the $\gamma_l\jpsi$ invariant mass distribution together
with the background estimated from the $\jpsi$ mass sidebands
(normalized to the width of the $\jpsi$ signal range) for the
combined $\EE$ and $\MM$ modes after the above selection criteria
are applied. Some ISR backgrounds with a correctly reconstructed $\jpsi$
remain in the data. No $\chicJ$ signal is observed.

A simultaneous fit to the signal region is performed with
Breit-Wigner (BW) functions convolved with Gaussian resolution
functions for the resonances and a second-order polynomial
background term. The width of the Gaussian resolution function is
fixed at $7.9$~MeV/$c^2$, which is obtained by increasing the
MC-simulated value by 10\% to account for the difference between
data and MC simulation. The masses and widths of the $\chicJ$
resonances are fixed to their PDG values~\cite{PDG}. In the
simultaneous fit, the ratio of the yields in the two $\jpsi$ decay
channels is fixed to $\BR_i\eff_i$, where $\BR_i$ is the
$\jpsi$ decay branching fraction for the $\EE$ mode or $\MM$ mode
reported by the PDG~\cite{PDG}, and $\eff_i$ is the MC-determined
efficiency for this mode. The upper limit on the number ($n^{\rm
up}$) of signal events at the 90\% C.L. is calculated by solving
the equation $\frac{\int_0^{n^{\rm up}}\LK(x)dx}
{\int_0^{+\infty}\LK(x)dx} = 0.9$, where $x$ is the number of
signal events, and $\LK(x)$ is the likelihood function depending
on $x$ from the fit to the data. The values of $n^{\rm up}$ are
found to be $2.8$, $3.1$ and $7.6$ for the $\chicz$, $\chico$ and
$\chict$, respectively,  when requiring the signal yields
to be non-negative in the fit. We do not observe any
structure at high masses, where  excited $\chicJ$ states are
expected.

%% gamma psip
To search for a possible excited charmonium state in the
$\gamma_l\psp$ final state, a $\jpsi$ candidate and two oppositely
charged pion candidates are reconstructed. The $\psp$ signal
region is defined as $3.67~\gevcs <M_{\ppjpsi}<3.70~\gevcs$, and
the $\psp$ mass sidebands are defined as $3.63~\gevcs
<M_{\ppjpsi}<3.66~\gevcs$ and $3.71~\gevcs < M_{\ppjpsi}<
3.74~\gevcs$. To suppress backgrounds with misconstructed photons,
we require the energy of the $\gamma_l$ to be higher than $75~\mev$.
To suppress the ISR background $\EE\to\gisr\psp\to\gisr\ppjpsi$,
we require the square of the mass recoiling against the $\gamma_l$
and $\psp$ to be within $-0.5~{\rm GeV}^2/c^4$ and 1.5~GeV$^2/c^4$
since $\MMS$ for the ISR background tends to be shifted towards 
negative values.

\begin{figure}[htbp]
\psfig{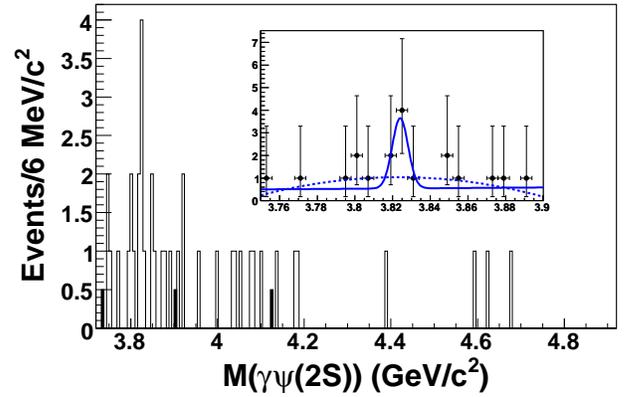} \caption{The
$\gamma_l\psp$ invariant mass distribution. The open histogram is
from the $\psp$ signal mass region, the shaded histogram is from
the normalized $\psp$ mass sidebands. In the inset, the solid
curve is the best fit between $3.75~\gevcs$ and $3.90~\gevcs$, and
the dashed curve is a fit with only a second-order polynomial
to describe the background. } \label{mgpsp}
\end{figure}

The ${\gamma_l\psp}$ invariant mass distribution after the above
selection is shown in Fig.~\ref{mgpsp}. There is no significant
signal. However, a few events accumulate around $3.82~\gevcs$,
where the $\gamma\psp$ decays of the $\chi_{c0}(2P)$ and
$\eta_{c2}(1D)$~\cite{Barnes} are expected. A fit between
$3.75~\gevcs$ and $3.90~\gevcs$ with a Gaussian to parameterize
the signal shape yields a mass of $(3.824\pm 0.002)~\gevcs$ and a
signal yield of $5.5\pm 2.7$ events corresponding to a
statistical significance of $1.8\sigma$. The signal significance
is determined by comparing the value of $-2\ln(L_0/L_{\rm max})$
from the fit, with values from fits to 10,000 pseudo-experiments.
Here $L_0$ and $L_{\rm max}$ are the likelihoods of the fits
without and with the signal, respectively. The upper limit on the
product branching fraction $\BR(\ytwos\to \gamma
X)\times\BR(X\to\gamma\psp)< 1.3 \times 10^{-5}$ at the 90\% C.L.
is determined following the procedure described below.

%% data selection for etac

To search for the $\etac$ signal in $\ytwos$ radiative decays, we
reconstruct $\etac$ candidates from the $\kkpi$, $\pp\kk$,
$2(\kk)$, $2(\pp)$, and $3(\pp)$ modes. Well measured charged
tracks should be identified as pions or kaons, and the number of
charged tracks is six for the $3(\pp)$ final state and four for
the other final states. In the $\kkpi$ mode, $K_S^0$ candidates
are reconstructed from $\pp$ pairs with an invariant mass
$M_{\pp}$ within 30~MeV/$c^2$ of the $K^0_S$ nominal mass. A
$K^0_S$ candidate should have a displaced vertex and flight
direction consistent with a $K^0_S$ originating from the IP; the
same selection method is used in Ref.~\cite{ks}. Events with
leptons misidentified as pions in the $\pp \kk$ and $2(\pp)$ modes
are removed by requiring $\mathcal{R}_e<0.9$ and
$\mathcal{R}_\mu<0.9$ for the pion candidates.  The value of
$\MMS$ for the hadronic daughters of the $\etac$ candidate is
required to be within $-1$~GeV$^2/c^4$ and 1~GeV$^2/c^4$.

\begin{figure}[htb]
\centerline{\psfig{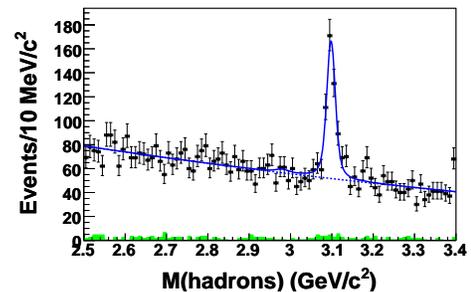}} \caption{
The mass distribution for a sum of the five $\etac$ decay modes.
The solid curve is a sum of the corresponding functions obtained
from a simultaneous fit to all the $\etac$ decay modes, and the
dashed curve is a sum of the background functions from the fit.
The shaded histogram is a sum of the off-resonance events (not
normalized). The $\jpsi$ signal is produced via ISR rather than
from a radiative decay of an $\Upsilon(nS)$ resonance.}
\label{metac}
\end{figure}

After the selection described above, Fig.~\ref{metac} shows the
combined mass distribution of the hadronic final states for the
five $\etac$ decay modes. The large $\jpsi$ signal is due to the
ISR process $\EE \to \gisr \jpsi$, while the accumulation of
events within the $\etac$ mass region is small. The shaded
histogram in Fig.~\ref{metac} is the same distribution for the
off-resonance data and is not normalized.

A simultaneous fit is performed to the five final states.
The ratios of the $\etac$ ($\jpsi$)
yields in all the channels are fixed to $\BR_i\eff_i$, where each
$\BR_i$ is the $\etac$ ($\jpsi$) decay branching fraction
for the $i$-th mode reported by the PDG~\cite{PDG}, and $\eff_i$
is the MC-determined efficiency for this mode. The fit function
contains a BW function convolved with a Gaussian resolution
function (its resolution is fixed to 7.9~$\hbox{MeV}/c^2$ from MC
simulation) describing the $\etac$ signal shape, another Gaussian
function describing the $\jpsi$ signal shape, and a second-order
polynomial describing the background shape. The mass and width of
the BW function are fixed to the PDG values~\cite{PDG} for the
$\etac$. The results of the fit are shown in Fig.~\ref{metac}, where
the solid curve is the sum of all the fit functions, and the
dashed curve is the sum of the background functions. The fit
yields $14\pm 20$ $\etac$ signal events
%, with a statistical
%significance of $0.7\sigma$. The
corresponding to an upper limit $n^{\rm up}$  of 44 at the 90\% C.L.
%for the number of $\etac$ signal events is estimated to be 44 at the 90\% C.L.
In addition, we obtain $370\pm
15$ $\jpsi$ signal events from the fit (in agreement with
$338\pm 16$ expected from $\gisr\jpsi$ production according to MC
simulation), giving a mass of $3098.1\pm 0.7~\hbox{MeV}/c^2$,
which is consistent with the PDG value~\cite{PDG}.

%% data selection for X(3872)

The selection criteria for $\ytwos\to \gR\x$, $\x\to \ppjpsi$ are
similar to those used for ISR $\ppjpsi$ events in $\Upsilon(4S)$
data~\cite{belley}.  We require that one $\jpsi$ candidate be
reconstructed, two well-identified $\pi$'s have an invariant mass
greater than 0.35~GeV/$c^2$, and that $\MMS(\ppjpsi)$ be within
the range between $-1$~GeV$^2/c^4$ and 1~GeV$^2/c^4$. To suppress
the ISR $\ppjpsi$ background, we require that the polar angle of
the $\gR$ candidate satisfy $|\cos\theta|<0.9$ in the $\EE$ C.M.
frame. Except for a few residual ISR produced $\psp$ signal
events, only a small number of events appear in the $\pp\jpsi$
invariant mass distribution, as shown in Fig.~\ref{x3872}(a).
There is no accumulation of events in the $\x$ mass region.
Fitting using a signal shape from the MC sample and a first-order
polynomial function as the background shape, the upper limit
$n^{\rm up}$ for the number of signal events is determined to be
3.6 at the 90\% C.L.

\begin{figure}[htbp]
\psfig{file=fig4a.epsi, width=6cm, angle=0}
\psfig{file=fig4b.epsi, width=6cm, angle=0}
\psfig{file=fig4c.epsi, width=6cm, angle=0} \caption{(a)
Distribution of the $\ppjpsi$ invariant mass for $\ytwos\to
\gR\ppjpsi$ candidates. (b) Distribution of the $\pp\piz\jpsi$
invariant mass for $\ytwos\to \gR\pp\piz\jpsi$ candidates. (c)
Scatter plots of $m(\pp \piz \jpsi)$ versus $m(\pp \piz)$, where
the region indicated by the ellipse corresponds to the
$\pm3\sigma$ mass regions of $m(\pp \piz \jpsi)$ and $m(\pp \piz)$
from the $X(3915)\to\omega\jpsi$ decay. Points with error bars are
data, open histograms are the MC expectation for the $X(3872)$
signal (arbitrary normalization). The peak at $3.686$~GeV/$c^2$ in
(a) is due to $\psp$ production via ISR.} \label{x3872}
\end{figure}

We also search for the $\x$ and $X(3915)$ in the $\pp\piz\jpsi$
mode. We select $\pi^+$, $\pi^-$, and $\jpsi$ candidates in the
$\x\to\ppjpsi$ mode (with the requirement on the $\pp$ invariant
mass greater than 0.35~GeV/$c^2$ removed) and a $\pi^0$ candidate
from a pair of photons with invariant mass within 10~MeV$/c^2$ of
the $\piz$ nominal mass. Here the $\piz$ mass resolution is about
4~MeV/$c^2$ from MC simulation. Figure~\ref{x3872}(b) shows the
$\pp\piz\jpsi$ invariant mass distribution, where the open
histogram is the MC expectation for the $X(3872)$ signal plotted
with an arbitrary normalization. Using the same fit method as in
$\x \to \pp \jpsi$, we determine $n^{\rm up}$ for the number of
$\x$ signal events to be 4.2 at the 90\% C.L.
Figure~\ref{x3872}(c) shows the scatter plot of $m(\pp \piz
\jpsi)$ versus $m(\pp \piz)$ from data, where the region indicated
by the ellipse corresponds to the $\pm3\sigma$ mass regions of
$m(\pp \piz \jpsi)$ and $m(\pp \piz)$ from the
$X(3915)\to\omega\jpsi$ decay. There is one event with $m(\pp \piz
\jpsi)$ at 3.923~GeV/$c^2$ and $m(\pp \piz)$ at 0.790~GeV/$c^2$
from $\Upsilon(2S)$ data, as shown in the ellipse. Assuming that
the number of background events is zero, the upper limit $n^{\rm
up}$ for the number of $X(3915)$ signal events is 4.4 at the 90\%
C.L.

We search for the $Y(4140)$ and the $X(4350)$ in the $\phi\jpsi$
mode. The selection criteria are very similar to those in the
analysis of $X(3872)\to \ppjpsi$ described above and the $\phi$ is
reconstructed from a $\kk$ pair. According to MC simulation, the
$\phi$ signal region is defined as
$1.01~\hbox{GeV}/c^2<M_{\kk}<1.03$~GeV/$c^2$. The number of well
measured charged tracks is required to be exactly four. After
applying all of the above event selection criteria, there is no
clear $\jpsi$ or $\phi$ signal. Nor are there candidate events in
the $Y(4140)$ or $X(4350)$ mass regions.  The upper limits on the
number of $Y(4140)$ and $X(4350)$ signal events are both 2.3 at
the 90\% C.L.

Several sources of systematic uncertainties are considered. The
uncertainty due to particle identification efficiency is
2.4\%-3.4\% and depends on the final state particles. The
uncertainty in the tracking efficiency for tracks with angles and
momenta characteristic of signal events is about 0.35\% per track,
and is additive. The photon reconstruction contributes an
additional 2.0\% per photon. Errors on the branching fractions of
the intermediate states are taken from the PDG~\cite{PDG}; they
are 6.9\% for the $\chicz$ mode, 4.5\% for the $\chico$ mode,
4.2\% for the $\chict$ mode, 1.7\% for the $\gamma\psp$ mode, 17\% for the $\eta_c$ mode,
 1.0\% for the $\x$ mode, 1.3\% for the
$X(3915)$ mode, and 1.6\% for the $\phi\jpsi$ mode. By using a
phase space distribution and including possible intermediate
resonant states, the largest difference of efficiency is
determined to be 2.1\% for the $\etac$ decay modes. The difference
in the overall efficiency for a flat angular distribution of
radiative photons and a $1\pm \cos^2\theta$ distribution is less
than 3.0\%. Therefore, we quote an additional error of 5.0\% due
to the limited knowledge of the decay dynamics for all the states
studied, except for the $\chi_{c0}$ mode and $\eta_c$ mode, which
are known to follow a $1+\cos^2\theta$ distribution. According to MC
simulation, the trigger efficiency is 89\% for the $\chicJ$ mode,
rather high for other modes ($\ge 99\%$); we take a 3.0\% error
for the $\chicJ$ mode and 1.0\% error for other modes as a
conservative estimate of the corresponding uncertainties. With the
pure $\EE\to\gisr\psp, \psp\to\ppjpsi$ or
$\jpsi\eta$($\to\gamma\gamma$) samples obtained from Belle data,
the uncertainty due to the recoil mass squared requirement is
1.0\% for the channels with a single photon and 4.7\% for channels
with two photons. By changing the order of the background
polynomial, the range of the fit, and the values of the masses and
widths of the resonances, uncertainties on the $\chicJ$ and
$\etac$ signal yields are estimated to be 1.1\% and 16\%,
respectively. In the $\ytwos \to \gR \chicJ$ mode, the uncertainty
associated with the requirement on the number of photons is 2.0\%
after applying a correction factor of 0.94 to the MC efficiency,
which is determined from a study of a very pure $\ytwos \to \MM$
event sample. In the $\etac \to \kkpi$ mode, the uncertainty in
the $K_S^0$ selection efficiency is determined by a study on a
large sample of high momentum $K_S^0 \to \pp$ decays; the
efficiency difference between data and MC simulation is less than
4.9\%~\cite{ks-error}. Finally, the uncertainty on the total
number of $\ytwos$ events is 2.3\%. Assuming that all of these
systematic error sources are independent, we add them in
quadrature to obtain a total systematic error as shown in
Table~\ref{summary}.

Since there is no evidence for signals in the modes studied, we
determine upper limits on the branching fractions of $\ytwos$
radiative decays. Table~\ref{summary} lists the upper limits
$n^{\rm up}$ for the number of signal events, detection
efficiencies, systematic errors, and final results for the upper
limits on the branching fractions. In order to calculate
conservative upper limits on these branching fractions, the
efficiencies are lowered by a factor of $1-\sigma_{\rm sys}$ in
the calculation.
%Most of the \mkred{ product branching fraction} upper limits are at the $10^{-6}$ level.

\begin{table}[htbp]
\caption{Summary of the limits on $\ytwos$ radiative decays to
charmonium and charmonium-like states $R$. Here $n^{\rm up}$ is
the upper limit on the number of signal events, $\eff$ is the
efficiency with the secondary decay branching fractions excluded
and trigger efficiency included, $\sigma_{\rm sys}$ is the total
systematic error, and $\BR(\ytwos \to \gamma R)^{\rm up}$ (${\cal
B}_R$) is the upper limit at the 90\% C.L. on the decay branching
fraction in the charmonium state case, and on the product
branching fraction in the case of a charmonium-like state.}
\label{summary}
\begin{center}
\begin{tabular}{c  c  c  c  c}
\hline
 State ($R$) & $n^{\rm up}$ &
$\eff$(\%) & $\sigma_{\rm
sys}$(\%)& ${\cal B}_R $ \\
\hline
 $\chicz$                    &  2.8  & 14.2 & 10.9 & $1.0\times 10^{-4}$ \\
 $\chico$                    &  3.1  & 14.8 &  10.8  & $3.6\times 10^{-6}$ \\
 $\chict$                    &  7.6  & 15.2 & 10.7  & $1.5\times 10^{-5}$ \\
 $\etac$                     &  44   & 26.3  & 24 & $2.7\times 10^{-5}$\\
% $X(3824) \to\gamma \psp$    &  10.6 & 14.4 & 9.6 & $1.3\times 10^{-5}$\\
 $\x\to\pi^+\pi^-\jpsi$      &  3.6  & 27.3 & 7.4  & $0.8\times 10^{-6}$\\
 $\x\to\pi^+\pi^-\pi^0 \jpsi$     &  4.2 & 10.3  & 9.6 & $2.4\times 10^{-6}$ \\
 $X(3915)\to\omega \jpsi$         &  4.4 & 10.5  & 9.6  & $2.8\times 10^{-6}$ \\
 $Y(4140) \to \phi\jpsi$          &  2.3 & 22.3  &   7.4  & $1.2\times 10^{-6}$\\
 $X(4350) \to \phi\jpsi $         &  2.3 & 21.0  &   7.4  & $1.3\times 10^{-6}$\\\hline
\end{tabular}
\end{center}
\end{table}

To summarize, we find no significant signals for the $\chicJ$ or
$\eta_c$, as well as for the $\x$, $X(3915)$, $\cdfy$, or
$X(4350)$ in $\ytwos$ radiative decays.
%In addition, we find no evidence for
%excited charmonium states in the invariant mass distributions of all final states
%below 5.5~GeV/$c^2$.
The results obtained on the $\chi_{cJ}$ and $\etac$ production
rates are consistent with the theoretical predictions
of~\cite{ktchao}.

%%%%%%%%%%%%%%%%%%%%%%%%%%%%%%%%%%%%%%%%%%%%%%%%%%%%%%%%%%%%%%%%
%%%%%    acknowledgments       Part                %%%%%%%%%%%%%
%%%%%%%%%%%%%%%%%%%%%%%%%%%%%%%%%%%%%%%%%%%%%%%%%%%%%%%%%%%%%%%%
We thank the KEKB group for excellent operation of the
accelerator, the KEK cryogenics group for efficient solenoid
operations, and the KEK computer group and the NII for valuable
computing and SINET4 network support. We acknowledge support from
MEXT, JSPS and Nagoya's TLPRC (Japan); ARC and DIISR (Australia);
NSFC (China); MSMT (Czechia); DST (India); MEST, NRF, NSDC of
KISTI, and WCU (Korea); MNiSW (Poland); MES and RFAAE (Russia);
ARRS (Slovenia); SNSF (Switzerland); NSC and MOE (Taiwan); and DOE
(USA).

\end{document}